\documentclass[twocolumn,floatfix,preprintnumbers,nofootinbib,superscriptaddress]{revtex4}

\usepackage{ulem}
\usepackage{bm}
\usepackage{times}
\usepackage{amssymb,amsbsy,amsmath,amsfonts}
\usepackage{graphicx}
\usepackage{float}
\usepackage{color}
\usepackage{morefloats}
\usepackage{rotating}
\usepackage{srcltx}
\usepackage{slashed}
\usepackage{subfigure}
\usepackage{multirow}
\usepackage{verbatim}
\usepackage{hyperref}
\usepackage{tabularx}

\usepackage[outdir=./]{epstopdf}

\begin{document}

\title{Faddeev fixed-center approximation to the $D\bar{D}K$ system and the hidden charm $K_{c\bar{c}}(4180)$ state}

\date{\today}

\author{Xiang Wei}~\email{weixiang@ucas.ac.cn}
\affiliation{Institute of Modern Physics, Chinese Academy of
Sciences, Lanzhou 730000, China} \affiliation{School of Nuclear
Sciences and Technology, University of Chinese Academy of Sciences,
Beijing 101408, China}

\author{Qing-Hua Shen}~\email{shenqinghua21@ucas.ac.cn}
\affiliation{Institute of Modern Physics, Chinese Academy of
Sciences, Lanzhou 730000, China} \affiliation{School of Nuclear
Sciences and Technology, University of Chinese Academy of Sciences,
Beijing 101408, China}

\author{Ju-Jun Xie}~\email{xiejujun@impcas.ac.cn}
\affiliation{Institute of Modern Physics, Chinese Academy of
Sciences, Lanzhou 730000, China} \affiliation{School of Nuclear
Sciences and Technology, University of Chinese Academy of Sciences,
Beijing 101408, China} \affiliation{School of Physics and
Microelectronics, Zhengzhou University, Zhengzhou, Henan 450001,
China}

\begin{abstract}

We perform a theoretical study on the $D\bar{D}K$ three body system, using the fixed center approximation to the Faddeev equations, considering the interaction between $D$ and $K$, $D$ and $\bar{D}$ from the chiral unitary approach. We assume the scattering of $K$ meson on a clusterized system $D\bar{D}$, where a scalar meson $X(3720)$ could be formed. Thanks to the strong $DK$ interaction, where the scalar $D^*_{s0}(2317)$ meson is dynamically generated, a resonance structure shows up in the modulus squared of the three body $K$-$(D\bar{D})_{X(3720)}$ scattering amplitude and supports that a $D\bar{D}K$ bound state can be formed. The result is in agreement with previous theoretical studies, which claim a new excited hidden charm $K$ meson, $K_{c\bar{c}}(4180)$ with quantum numbers $I(J^P) = \frac{1}{2}(0^-)$ and mass about $4180$ MeV. It is expected that these theoretical results motivate its search in experimental measurements.

\end{abstract}

\maketitle

\section{Introduction} \label{sec:introduction}

The study of hadron structure and spectrum is one of the most important issues in hadronic physics and is attracting much attention. The traditional quark model classifies hadrons into the mesons which are composed of a pair of quark and anti-quark ($q\bar{q}$), and the baryons which are composed of three quarks ($qqq$). That well explains the basic properties of most of the experimentally discovered mesons and baryons~\cite{Godfrey:1985xj,Capstick:1986kw,Capstick:1986ter}. However, the existence of exotic states and the investigation of their properties will extend our knowledge of the
strong interaction dynamics~\cite{Chen:2016qju,Chen:2016spr,Guo:2017jvc,Chen:2022asf}. Recently, the discovery of the $Z_{cs}(3985)$ in the $D^{*-}_s D^0$ and $D^-_s D^{*0}$invariant mass distribution of the $e^+e^- \to K^+ (D^{*-}_s D^0 + D^-_s D^{*0})$ reaction by the BESIII collaboration in Ref.~\cite{BESIII:2020qkh} was a turning point in hadron spectroscopy since it provided a clear example of an exotic meson state. However, its quantum numbers need to be determined by more precise experimental data. The $Z_{cs}(3985)$ has the same valence quarks with the $K$ meson, and it could be the first candidate for a charged hidden-charm tetraquark with strangeness. On the theoretical side, there are much work that take the $Z_{cs}(3985)$ as a molecular state by using directly dynamics for meson-meson interactions~\cite{Hidalgo-Duque:2012rqv,Wu:2021ezz,Ding:2021igr,Meng:2021rdg,Ikeno:2020mra,Wang:2020htx,Sun:2020hjw,Du:2020vwb,Guo:2020vmu,Yan:2021tcp,Ikeno:2021mcb,Du:2022jjv,Baru:2021ddn,Zhai:2022ied}.

Meanwhile, there is also evidence that some excited hadronic states are predicted in terms of bound state of resonances of three particles. For example, the $D\bar{D}K$ three body system, with hidden charm, was investigated in Ref.~\cite{Wu:2020job} by solving the Schr\"odinger equation with the Gaussian Expansion Method. Thanks to the strong interaction of $DK$, it was found that the $D\bar{D}K$ system can form a bound state, the excited $K$ meson, with quantum numbers $I(J^P) = \frac{1}{2}(0^-)$ and with mass about $4180$ MeV. Indeed, the effective $DK$ interaction is strong. Within the chiral dynamics the scalar $D^*_{s0}(2317)$ meson could be explained as a bound $DK$ state~\cite{Kolomeitsev:2003ac,Guo:2006fu,Gamermann:2006nm,Geng:2010vw,Mohler:2013rwa,Altenbuchinger:2013vwa,Guo:2015dha,Huang:2022cag}.

For the $D\bar{D}$ interaction, a scalar $D\bar{D}$ bound state was predicted in Ref.~\cite{Zhang:2006ix} by solving the Schr\"odinger equation with one vector meson exchange potential. Within the chiral unitary approach, the scalar $D\bar{D}$ bound state $X(3720)$ was obtained with mass about $3720$ MeV~\cite{Gamermann:2006nm,Xiao:2012iq}. This $D\bar{D}$ state was studied in the $B^0 \to D^0 \bar{D}^0K^0$ and $B^+ \to D^0 \bar{D}^0 K^+$ reactions in Refs.~\cite{Dai:2015bcc}, $\psi(3770) \to \gamma D^0\bar{D}^0$ reaction~\cite{Dai:2020yfu}, and $\Lambda_b \to \Lambda D\bar{D}$ reaction~\cite{Wei:2021usz}. Based on the experimental measurements, the analyses of  $e^+ e^- \to J/\psi D\bar{D}$ and $\gamma \gamma \to D \bar{D}$ reactions were done in Refs.~\cite{Wang:2020elp,Wang:2019evy}, where the evidence of the existence of a $S$-wave $D\bar{D}$ bound state was also claimed. Recently, a scalar $D\bar{D}$ bound state was also found according to the lattice calculation in Ref.~\cite{Prelovsek:2020eiw}. Furthermore, in Refs.~\cite{LHCb:2020bls,LHCb:2020pxc}, the contributions from $S$-wave $D^+D^-$ interactions were also investigated in the $B^+ \to D^+D^-K^+$ decay by the LHCb collaboration.

Motivated by the work of Ref.~\cite{Wu:2020job}, we reinvestigate the three body $D\bar{D}K$ system by considering the strong interactions of $D\bar{D}$ and $DK$. Making the $D\bar{D}$ bound state, $X(3720)$, as a cluster, and in terms of the two body $DK$ and $\bar{D}K$ scattering amplitudes obtained within the chiral unitary approach, we solve the Faddeev equations by using the fixed center approximation (FCA). The main purpose of this work is to test the validity of the FCA to study the $D\bar{D}K$ system which was done in Ref.~\cite{Wu:2020job} with the Gaussian expansion method.

The FCA has been employed before, in particular in the similar systems to the $D\bar{D}K$. For example: the $DKK$ and $DK\bar{K}$ systems~\cite{Debastiani:2017vhv}; the $D\bar{D}^*K$ system~\cite{Ren:2018pcd}, the $BD\bar{D}$ and $BDD$ systems~\cite{Dias:2017miz}. In the baryon sector, within the FCA, the $NDK$ and $ND \bar{D}$ systems were studied in Ref.~\cite{Xiao:2011rc}, while the $ND\bar{D}^*$ system was studied in Ref.~\cite{Malabarba:2021taj}. Besides, in Ref.~\cite{Ma:2017ery}, the Schr\"odinger equation for the $D\bar{D}^*K$ system was solved, and very similar results were obtained in Ref.~\cite{Ren:2018pcd}. In Ref.~\cite{MartinezTorres:2018zbl}, by solving the full Faddeev equations of the $DDK$ system considering the $DD_s\eta$ and $DD_s \pi$ coupled channels, the three-body scattering amplitudes were obtained. It was found that an isospin $I=1/2$ state is formed at 4140 MeV, which is compatible with the one found in Ref.~\cite{SanchezSanchez:2017xtl} where the $D$-$D^*_{s0}(2317)$ system was studied without considering explicit three-body dynamics.

This may indicate that the FCA is a good approximation for the problem where one has a cluster of two particles and allows the third particle to undergo multiple scattering with this cluster. And the cluster is assumed not to be changed by the interaction with the third particle. Intuitively this would mostly happen when the third particle has a smaller mass than the particles in the cluster~\cite{Kamalov:2000iy,Xie:2010ig,Xie:2011uw}. This is just the situation of the present $D\bar{D}K$ system, where we keep the strong correlations of the $D\bar{D}$ system that generate the scalar meson $X(3720)$.

Along this line, in this work, we mainly study the $D\bar{D}K$ three body system by using the FCA approach, and prove it can produce a bound state. On the other hand, we take the advantage that the $D\bar{D}K$ system was already studied with other approach in Ref.~\cite{Wu:2020job}, such that comparison with the result of the above work can give us a feeling of the accuracy of the FCA, which is, technically, much easier than most of other methods to solve the full Faddeev equations. 

The paper is organized as follows. In Sec.~\ref{sec:formalism}, we present the theoretical formalism for studying the $D\bar{D}K$ system with the FCA, and in Sec.~\ref{sec:results}, we show our numerical results. Finally, a short summary is given in the last section.

\section{Formalism} \label{sec:formalism}

We are going to use the FCA of the Faddeev equations in order to obtain the scattering amplitude of the three body $D\bar{D}K$ system, where $D\bar{D}$ is considered as a bound state of $X(3720)$, which allows us to use the FCA to solve the Faddeev equations. From the analysis of the $KX(3720) \to KX(3720)$ scattering amplitude one could study the dynamically generated hidden charm $K_{c\bar{c}}$ states.

The important ingredients in the calculation of the total scattering amplitude for the $D\bar{D}K$ system using the FCA are the two-body $D\bar{D}$ and $DK$ unitarized s-wave interactions from the chiral unitary approach. Since the form of these two body interactions have been reported in many previous works, we direct the reader for details to Refs.~\cite{Gamermann:2006nm,Geng:2010vw,Mohler:2013rwa}, thus we will directly start with the form factor of $X(3720)$ that is a bound state of $D\bar{D}$.

\subsection{Form factor for the $X(3720)$} \label{subsec:formfactor}

\begin{figure}[htbp]
\centering
\includegraphics[scale=0.35]{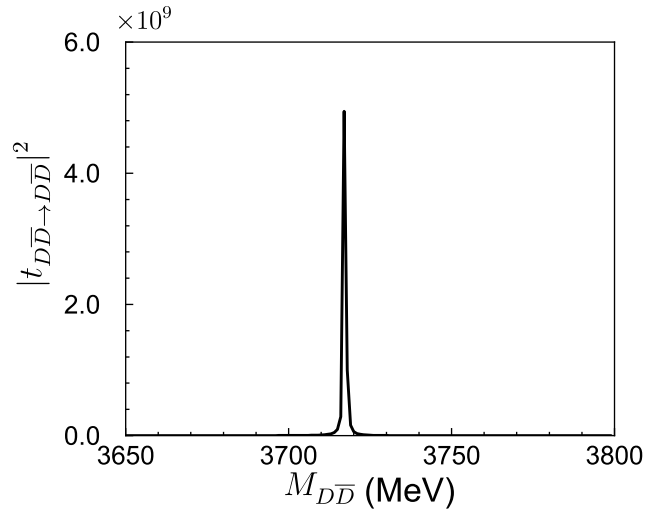}
\caption{Modulus squared of $t_{D\bar{D} \to D\bar{D}}$ in the sector of $I=0$ as a function of the invariant mass $M_{D\bar{D}}$ of the subsystem $D\bar{D}$.}
\label{fig:DDbarboundstate}
\end{figure}

We firstly redo the work of Ref.~\cite{Gamermann:2006nm} including the channels $D^+_sD^-_s$, $D\bar{D}$, $\eta \eta_c$, $\eta \eta$, $K\bar{K}$, and $\pi \pi$. Using a cutoff regularization method for the two body loop function and the transition potentials derived as in Ref.~~\cite{Gamermann:2006nm}, the Bethe-Salpeter equations can be solved, and then one gets the two body scattering amplitude $t_{D\bar{D}\to D\bar{D}}$ in the isospin $I=0$ sector. In Fig.~\ref{fig:DDbarboundstate} we show the modulus squared of the scattering amplitude $|t_{D\bar{D}\to D\bar{D}}|^2$ as a function of the invariant mass $M_{D\bar{D}}$ of the $D\bar{D}$ system, where the peak of $X(3720)$ state is clearly seen. Note that the numerical results are obtained with a cutoff $\Lambda = 850$ MeV to regularize the loop function for the integral of intermediate two meson propagators. It is found that, by adjusting the value of $\Lambda$, there is always a clear peak. For example, with $\Lambda = 950$ MeV, one could get a peak around $3700$ MeV. In addition, the masses of the particles considered in this work are shown in table~\ref{tab:particlemass}, which are taken from the review of particle physics~\cite{ParticleDataGroup:2020ssz}.

\begin{table}[htbp]
\renewcommand\arraystretch{1.5}
\caption{Particle masses (in MeV) used in this calculation.}
\begin{tabular}{c|c|c|c|c|c|c}
\hline
\hline
Particle & $D (\bar{D})$ & $D^+_s (D^-_s)$ & $K (\bar{K})$ & $\pi$ & $\eta$ & $\eta_c$ \\ \hline
Mass & 1869.12 & 1969.49 & 495.6455 & 138.04 & 547.862 & 2980.916 \\ \hline \hline
\end{tabular}
\label{tab:particlemass}
\end{table}

Next, following Refs.~\cite{Gamermann:2009uq,Yamagata-Sekihara:2010kpd}, one can easily obtain the expression of the form factor $F_R(q)$ for the bound state $X(3720)$, which is given by~\cite{Yamagata-Sekihara:2010kpd,Yamagata-Sekihara:2010muv,Roca:2010tf},
\begin{eqnarray}
F_R(q) &=& \frac{1}{\cal N} \int_{|\vec p~\!|<\Lambda,|\vec p - \vec q~\!|<\Lambda}d^3\vec p \frac{1}{4 E_1(\vec p~\!) E_2(\vec p~\!)} \nonumber \\
&& \times \frac{1}{M - E_1(\vec p~\!) - E_2(\vec p~\!)}\frac{1}{4 E_1(\vec p-\vec q~\!) E_2(\vec p-\vec q~\!)} \nonumber \\
&& \times \frac{1}{M - E_1(\vec p-\vec q~\!) - E_2(\vec p-\vec q~\!)},
\label{eq:formfactor}
\end{eqnarray}
with the normalization factor ${\cal N}$, which is
\begin{eqnarray}
{\cal N} =  \int_{|\vec p~\!|<\Lambda}d^3\vec p \left (\frac{1}{4 E_1(\vec p~\!) E_2(\vec p~\!)} \frac{1}{M - E_1(\vec p~\!) - E_2(\vec p~\!)} \right )^2,
\end{eqnarray}
where $M$ is the mass of $X(3720)$, $E_1$ and $E_2$ are the energies of $D$ and $\bar{D}$, respectively. The cutoff parameter $\Lambda$ is needed to regularize the loop functions in the chiral unitary approach.

\begin{figure}[htbp]
\centering
\includegraphics[scale=0.35]{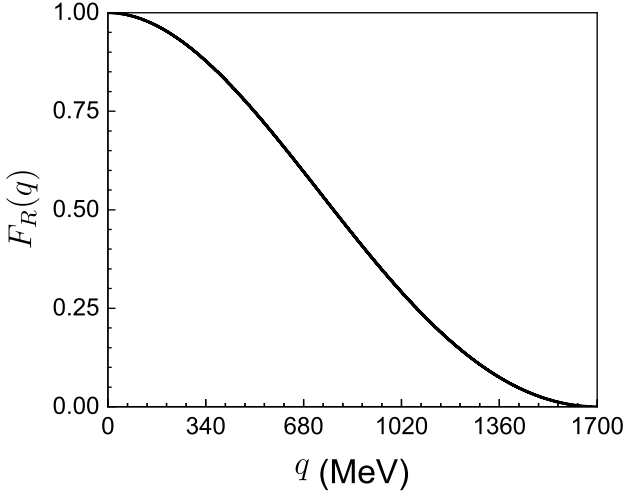}
\caption{Form factor for the $X(3720)$ as a $D\bar{D}$ bound state.} \label{fig:formfactor}
\end{figure}

In fig.~\ref{fig:formfactor} we show the results of the form factor for $X(3720)$ as a function of $q$, where we take $M= 3720$ MeV and $\Lambda = 850$ MeV. In Eq.~\eqref{eq:formfactor} the condition $|\vec{p} - \vec{q}~\!| < \Lambda$ implies that the form factor is exactly zero for $q > 2\Lambda$. Therefore the integration in Eq.~\eqref{eq:formfactor} has an upper limit of $2\Lambda$.

\subsection{Faddeev equations under fixed center approximation} \label{subsec:FCA}

\begin{figure*}[htbp]
\centering \includegraphics[scale=0.95]{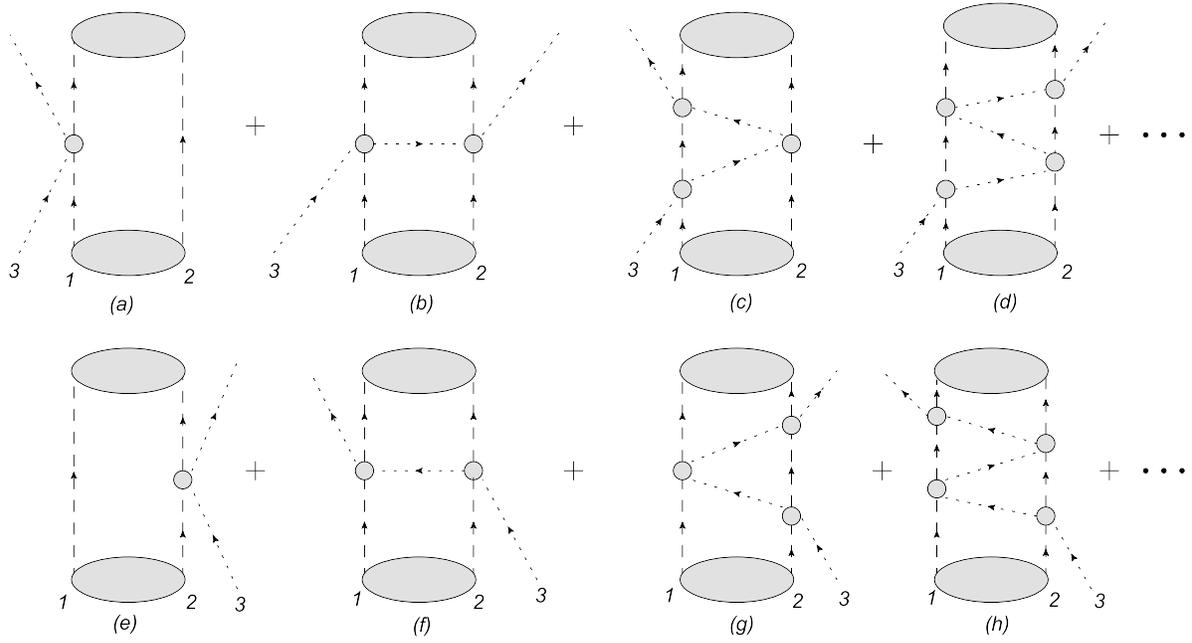}
\caption{Graphical representation of Faddeev equation under fixed center approximation.}
\label{fig:diagramsFCA}
\end{figure*}

In the framework of FCA, we consider the $D\bar{D}$ bound state $X(3720)$ as a cluster, and the $K$ meson interacts with the components of the cluster. The total three-body scattering amplitude $T$ can be simplified as the sum of two partition functions $T_1$ and $T_2$, by summing all diagrams in Fig.~\ref{fig:diagramsFCA}, starting from the interaction of particle 3 with particle 1(2) of the cluster. Thus, the FCA equations can be written in terms of $T_1$ and $T_2$, which read~\cite{Barrett:1999cw,Deloff:1999gc}
\begin{eqnarray}
T_1 &=& t_1 + t_1 G_0 T_2, \label{eq:FCAT1} \\
T_2 &=& t_2 + t_2 G_0 T_1, \label{eq:FCAT2} \\ 
T &=& T_1+T_2,  \label{eq:FCATtotal}
\end{eqnarray}
where $T_1$ represents the amplitudes of all multiple scattering processes in which particle 3 first scatters with particle 1 inside the cluster, similarly, $T_2$ represents the amplitudes of all multiple scattering processes in which particle 3 first scatters with particle 2 inside the cluster. While $t_1$ and $t_2$ represent, in the present work, the two body scattering amplitudes of $DK \to DK$ and $\bar{D}K \to \bar{D}K$, respectively. We will take the results in the work of Ref.~\cite{Gamermann:2006nm} for $t_1$ and $t_2$, which will be discussed in following.

In Eqs.~\eqref{eq:FCAT1} and \eqref{eq:FCAT2}, the $G_0$ is the $K$ meson propagator between the $D$ and $\bar{D}$ in the cluster, which is,
\begin{equation}
G_0(s)=\frac{1}{2M}\int \frac{d^3 \vec q}{(2\pi)^3} \frac{F_{R}(q)}{(q^0)^2 - E^2_K(q) + i\epsilon},
\label{eq:loopfunctionGzero}
\end{equation}
with $E^2_K(q) = |\vec{q}~\!|^2 + m^2_K$, and $q^0$ is the energy of $K$ meson in the rest frame of the cluster where the form factor $F_R(q)$ is calculated, and its expression is:
\begin{equation}
q^0(s) = \frac{s + m_K^2 - M^2}{2\sqrt{s}}
\label{eq:qzero}
\end{equation}
where $s$ is the invariant mass squared of the $D\bar{D}K$ three body system.

\begin{figure}[htbp]
\centering
\includegraphics[scale=0.35]{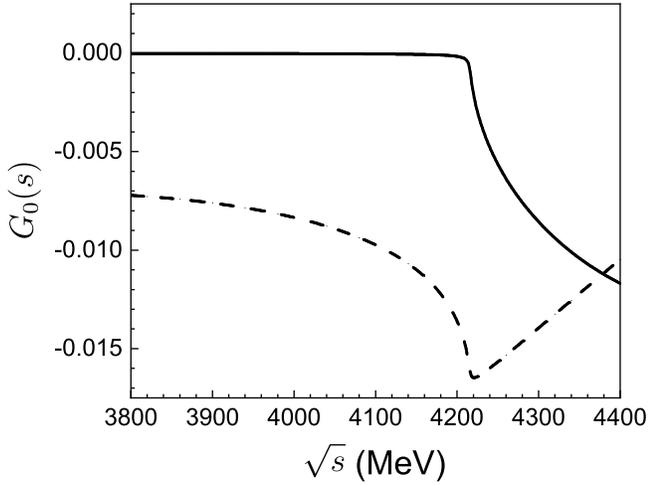}
\caption{Real (solid line) and imaginary (dash line) parts of $G_0$ as a function of the total energy of the three body $D\bar{D}$K system.} \label{fig:Gzero}
\end{figure}

In Fig.~\ref{fig:Gzero}, we show the real (red line) and imaginary
(blue line) parts of $G_0$ as a function of the invariant
mass of the $D\bar{D}K$ system.

\subsection{Scattering amplitude of the three body $D\bar{D}K$ system} \label{subsec:amplitudeTtotal}

In this work we study the $K(D\bar{D})_{X(3720)}$ configuration of the $KD\bar{D}$ system, which means that we need to study the $D\bar{D}K$ system where $D\bar{D}$ is treated as $X(3720)$. Since $D$ and $\bar{D}$ are isospin $1/2$ states, then the $D\bar{D}$ isospin $I=0$ state can be written as
\begin{equation}
|D\bar{D}\rangle_{I=0} = \frac{1}{\sqrt{2}}|(\frac{1}{2},-\frac{1}{2})\rangle - \frac{1}{\sqrt{2}}|(-\frac{1}{2},\frac{1}{2})\rangle ,
\label{eq:DDbarIzero}
\end{equation}
where the last numbers in the kets indicate the $I_z$ components of $D$ and $\bar{D}$ mesons, $(|I^z_D,I^z_{\bar{D}}\rangle)$. We take $D = (D^+, D^0)$ and $\bar{D} = (\bar{D}^0, D^-)$.

Then the three body scattering amplitude $\langle D\bar{D}K|t|D\bar{D}K\rangle$ can be easily obtained in terms of the two-body potentials $V_{KD}$ and $V_{K\bar{D}}$ derived in Refs.~\cite{Gamermann:2006nm,Xiao:2012iq}, which is written as:
\begin{widetext}
\begin{equation}
\begin{aligned}
T&=\langle D\bar{D}K|V|D\bar{D}K\rangle=\{ \langle \frac{1}{2}|\otimes \frac{1}{\sqrt{2}}(\langle \frac{1}{2},-\frac{1}{2}|-\langle -\frac{1}{2},\frac{1}{2}|)\} (V_{KD} + V_{K\bar{D}}) \{|\frac{1}{2}\rangle \otimes \frac{1}{\sqrt{2}}(|\frac{1}{2},-\frac{1}{2}\rangle-|-\frac{1}{2},\frac{1}{2}\rangle)\}\\
&=\frac{1}{2}\{ [\langle (1,1),-\frac{1}{2}|-\frac{1}{\sqrt{2}}\langle (1,0),\frac{1}{2}|+\frac{1}{\sqrt{2}}\langle (0,0),\frac{1}{2}|]V_{KD}[|(1,1),-\frac{1}{2}\rangle- \frac{1}{\sqrt{2}}|(1,0),\frac{1}{2}\rangle+\frac{1}{\sqrt{2}}|(0,0),\frac{1}{2}\rangle]\}\\
&+[\frac{1}{\sqrt{2}}\langle \frac{1}{2},(1,0)|-\frac{1}{\sqrt{2}}\langle \frac{1}{2},(0,0)|-\langle -\frac{1}{2},(1,1)|]V_{K\bar{D}}[ \frac{1}{\sqrt{2}}|\frac{1}{2},(1,0)\rangle-\frac{1}{\sqrt{2}}|\frac{1}{2},(0,0)\rangle-|-\frac{1}{2}\rangle,(1,1)],
\end{aligned}
\label{eq:totalT}
\end{equation}
\end{widetext}
where the notation for the states in the last equality is $[(I_{KD}I^z_{KD}),I^z_D]$ for $t_1$ and $[(I_{K\bar{D}}I^z_{K\bar{D}}),I^z_{\bar{D}}]$ for $t_2$. This leads to the following amplitudes for the single-scattering contribution,
\begin{eqnarray}
t_1 &=& \frac{3}{4}t_{DK}^{I=1}+\frac{1}{4}t_{DK}^{I=0}, \label{eq:t1} \\
t_2 &=& \frac{3}{4}t_{\bar{D}K}^{I=1}+\frac{1}{4}t_{\bar{D}K}^{I=0}. \label{eq:t2}
\end{eqnarray}

It is worth noting that the argument of the total scattering amplitude $T$ depends on the total invariant mass squared $s$, while the argument in $t_1$ is $s_{DK}$ and in $t_2$ is $s_{\bar{D}K}$, where $s_{DK}$ and $s_{\bar{D}K}$ are the invariant masses squared of the external $K$ meson with $D$ and $\bar{D}$ inside the cluster of $X(3720)$, respectively. In the rest frame of the $X(3720)$, $s_{DK}$ and $s_{\bar{D}K}$ are given by
\begin{eqnarray}
s_{DK} &=& m^2_K + m^2_D + \frac{s - m^2_K - M^2}{2}  ,\\
s_{\bar{D}K} &=& m^2_K + m^2_{\bar{D}} + \frac{s - m^2_K - M^2}{2} .
\end{eqnarray}

Before proceeding further, a normalization factor is needed for the two body scattering amplitudes $t_1$ and $t_2$,
\begin{eqnarray}
t_1 &\to& \frac{M}{m_D} t_1 , \\
t_2 &\to& \frac{M}{m_{\bar{D}}} t_2. 
\end{eqnarray}

With all the above ingredients, the total scattering amplitude of $D\bar{D}K$ three body system can be easily obtained,
\begin{equation}
T=\frac{t_1+t_2+2t_1t_2G_0}{1-t_1t_2G_0^2}.
\end{equation}
From the analysis of the $K(D\bar{D})_{X(3720)}$ scattering amplitude $T$ one can identify dynamically generated resonances with peaks in $|T|^2$.

\section{Numerical results and conclusions} \label{sec:results}

In this section we show the numerical results obtained for the $D\bar{D}K$ three body scattering amplitude squared with total isospin $I=1/2$ and spin-parity $J^P = 0^-$. We evaluate the scattering amplitude $T$ and search for peaks in $|T|^2$, which are identified with three body states generated from the $K(D\bar{D})_{X(3720)}$ system.

\begin{figure}[htbp]
\centering
\includegraphics[scale = 0.35]{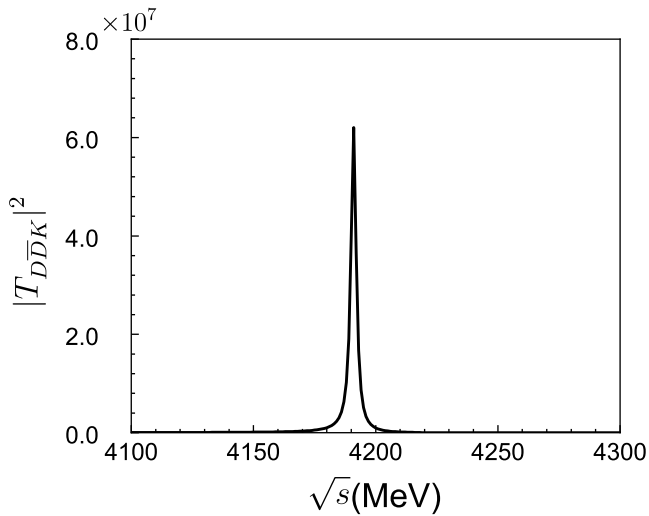}
\caption{Modulus squared of the total scattering amplitude of $D\bar{D}K$ three body system in isospin $I = 1/2$.} \label{fig:Tsquare}
\end{figure}

In fig.~\ref{fig:Tsquare} we show the modulus squared $|T|^2$ for the $KX(3720) \to KX(3720)$ scattering. One can see that there is a clear and narrow peak around $4191$ MeV, which could be associated with the state, $K_{c\bar{c}}(4180)$, obtained in Ref.~\cite{Wu:2020job}. However, it was found that the root-mean-square radius of the $DK$ subsystem in the $D\bar{D}K$ bound state is about $1.26$ fm, while those of the $\bar{D}K$ and $D\bar{D}$ subsystems are much larger, yielding $2.27$ and $2.10$ fm, respectively. Here, we take the $D\bar{D}$ subsystem as a bound state of $X(3720)$. Furthermore, taking $\sqrt{s} = 4191$ MeV, we get $\sqrt{s_{DK}} =\sqrt{s_{\bar{D}K}}= 2342$ MeV. At this energy, the interactions of $DK$ and $\bar{D}K$ are strong.

It is interesting to mention that a study of Ref.~\cite{Di:2019qwv}, within QCD sum rules, do not find a $\bar{D}D^*_{s0}(2317)$ bound state, which may indicate that the $K_{c\bar{c}}(4180)$ is a three body $D\bar{D}K$ molecule, where the $D\bar{D}$, $DK$, and $\bar{D}K$ interactions are important. Though, it is clear that the existence of a $D\bar{D}K$ bound state seems to be a robust prediction.

\section{Summary}

In summary, we have investigated the $D\bar{D}K$ three body system assuming that there is a primary clustering of particles $KX(3720)$ with $X(3720)$ as a bound state of $D\bar{D}$ subsystem. By using the fixed center approximation to the Faddeev equations, we have obtained the $KX(3720) \to KX(3720)$ scattering amplitude. It is found that there is a clear and narrow peak in $|T|^2$ around $4180$ MeV indicating the formation of a bound $D\bar{D}K$ state around this energy. These results are in agreement with those obtained in Ref.~\cite{Wu:2020job} using the Gaussian expansion method. Thus, the main value of the present work is not only to provide extra support for the existence of $K_{c\bar{c}}(4180)$ state but also to test the reliability of the FCA to deal with the $D\bar{D}K$ system.

Finally, we would like to stress that, thanks to the strong $s$-wave interactions of $D\bar{D}$ and $DK$, the $D\bar{D}K$ system can bind. This support the existence of a excited hidden charm $K$ meson, $K_{c\bar{c}}(4180)$, with quantum numbers $I(J^P) = \frac{1}{2}(0^-)$ and mass about $4180$ MeV. So far there is no experimental data available on this state~\cite{ParticleDataGroup:2020ssz}. It is expected that these theoretical results motivate its search in the future experimental measurements (see more discussions in Refs.~\cite{Wu:2019vsy,Wu:2020job,Wu:2021dwy}).

\begin{acknowledgments}

We would like to thank Prof. Li-Sheng Geng and Prof. En Wang for useful discussions. This work is partly supported by the National Natural Science Foundation of China under Grant Nos. 12075288, 11735003, and 11961141012. It is also supported by the Youth Innovation Promotion Association CAS.

\end{acknowledgments}

\bibliography{refence.bib}

\end{document}